\documentstyle[12pt]{article}
\newlength{\extraspace}
\newlength{\extraspaces}
\newcommand{\be}{\begin{equation}
\addtolength{\abovedisplayskip}{\extraspaces}
\addtolength{\belowdisplayskip}{\extraspaces}
\addtolength{\abovedisplayshortskip}{\extraspace}
\addtolength{\belowdisplayshortskip}{\extraspace}}
\newcommand{\ee}{\end{equation}}
\newcommand{\ba}{\begin{eqnarray}
\addtolength{\abovedisplayskip}{\extraspaces}
\addtolength{\belowdisplayskip}{\extraspaces}
\addtolength{\abovedisplayshortskip}{\extraspace}
\addtolength{\belowdisplayshortskip}{\extraspace}}
\newcommand{\ea}{\end{eqnarray}}
\newcommand{\nonu}{\nonumber \\[.5mm]}
\newcommand{\A}{&\!\!\!}
\begin{document}
\thispagestyle{empty}
\setlength{\baselineskip}{6mm}
\begin{flushright}
SIT-LP-09/12 \\
December, 2009
\end{flushright}
\vspace{7mm}
\begin{center}
{\large \bf Super Yang-Mills theory from nonlinear supersymmetry
} \\[20mm]
{\sc Kazunari Shima}
\footnote{
\tt e-mail: shima@sit.ac.jp} \ 
and \ 
{\sc Motomu Tsuda}
\footnote{
\tt e-mail: tsuda@sit.ac.jp} 
\\[5mm]
{\it Laboratory of Physics, 
Saitama Institute of Technology \\
Fukaya, Saitama 369-0293, Japan} \\[20mm]
\begin{abstract}
The relation between a nonlinear supersymmetic (NLSUSY) theory 
and a SUSY Yang-Mills (SYM) theory is studied for $N = 3$ SUSY in two-dimensional space-time. 
We explicitly show the NL/L SUSY relation for the (pure) SYM theory 
by means of cancellations among Nambu-Goldstone fermion self -interaction terms. 
\\[5mm]
\end{abstract}
\end{center}

\newpage

\noindent
Various linear supersymmeric (LSUSY) theories \cite{WZ1,WZ2,GL} are related to the Volkov-Akulov (VA) model \cite{VA} 
for nonlinear representation of supersymmetry (NLSUSY). 
The relation (abbreviated as NL/L SUSY relation) gives new insights 
into the low energy physics based on the nonlinear supersymmetric general relativity (NLSUSY GR) theory \cite{KS1} 
in SGM scenario \cite{KS1}-\cite{ST3-SGM} 
towards the unified description of space-time, all forces and matter beyond the standard model (SM). 

The linearization of NLSUSY \cite{IK1}-\cite{UZ} in flat space-time was shown 
in the various LSUSY (free) theories with the spontaneous SUSY breaking (SSB) \cite{FI} 
for $N = 1$ SUSY \cite{IK1}-\cite{STT1} and for extended SUSY, i.e. $N = 2$ \cite{STT2} 
and $N = 3$ SUSY (in two-dimensional space-time ($d = 2$)) \cite{ST4-lin}, 
and also shown in interacting $N = 2$ LSUSY (Yukawa interaction and SUSY QED) theories 
\cite{ST5-lin}-\cite{ST8-lin} in $d = 2$. 
(Note that for $N = 2$ SUSY in SGM scenario $J^P = 1^-$ gauge field appears \cite{STT2} in the linearization of NLSUSY. 
Therefore $N = 2$ SUSY is realistic minimal case.) 
%
By using the NL/L SUSY relation for the $N = 2$ SUSY QED theory (in $d = 2$) we have shown \cite{ST9,STL} 
that LSUSY-supermultiplet fields are realized as the composite (massless) eigenstates of the spin-${1 \over 2}$ 
masless Nambu-Goldstone (NG) fermions ({\it superons}) in the (fundamental) NLSUSY theory 
on the true vacuum (the minimum of the potential). 
As for the physical significances of NL/L SUSY relation, 
we pointed out that since the SSB scale originated from the cosmological term in NLSUSY GR 
induces a fundamental mass scale on the true vacuum, 
it gives a natural explanation of the mysterious (observed) numerical relation 
between the (four dimensional) dark energy density of the universe and the neutrino ($\nu$) mass \cite{ST3-SGM,ST9,STL} 
in the asymptotically flat space-time, provided $\nu$ is a composite of superons of these kinds. 

Beyond the abovementioned works for the linearization of NLSUSY, 
the extension of NL/L SUSY relation to non-Abelian gauge theories is a crucial and interesting problem. 
In this letter we address the problem by focusing on $N = 3$ (pure) super Yang-Mills (SYM) theory in $d = 2$ 
for simplicity of calculations without the loss of generalities. 
SUSY invariant relations which connect LSUSY theories with the NLSUSY model 
play an important role for establishing the NL/L SUSY relation, 
where basic fields in LSUSY supermultiplets are expressed as the composites of superons in a SUSY invariant way. 
We construct the SUSY invariant relations for the SYM theory 
and explicitly discuss the relation between the $N = 3$ NLSUSY action and a $N = 3$ (pure) SYM action. 
Each interaction terms in the $d = 2$, $N = 3$ SYM theory does not vanish in terms of superons, 
but the NL/L SUSY relation is realized by means of (nontrivial) cancellations among those terms as discussed later. 

Let us first introduce the VA NLSUSY action for $N$ SUSY \cite{VA,BV}, 
\be
S_{\rm NLSUSY} = - {1 \over {2 \kappa^2}} \int d^2x \ \vert w \vert, 
\label{NLSUSYaction}
\ee
where $\kappa$ is a constant whose dimension is (mass)$^{-1}$ 
and $\vert w \vert$ is the determinant describing the dynamics 
of (Majorana) superons $\psi^i(x)$ ($i, j, \cdots = 1, \cdots, N$), 
which is written in $d = 2$ 
\footnote{
Minkowski space-time indices in $d = 2$ are denoted by $a, b, \cdots = 0, 1$. 
The Minkowski space-time metric is 
${1 \over 2}\{ \gamma^a, \gamma^b \} = \eta^{ab} = {\rm diag}(+, -)$ 
and $\sigma^{ab} = {i \over 2}[\gamma^a, \gamma^b] 
= i \epsilon^{ab} \gamma_5$ $(\epsilon^{01} = 1 = - \epsilon_{01})$, 
where we use the $\gamma$ matrices defined as $\gamma^0 = \sigma^2$, 
$\gamma^1 = i \sigma^1$, $\gamma_5 = \gamma^0 \gamma^1 = \sigma^3$ 
with $(\sigma^1, \sigma^2, \sigma^3)$ being Pauli matrices. 
}
as 
\be
\vert w \vert = \det(w^a{}_b) = \det(\delta^a_b + t^a{}_b) 
= 1 + t^a{}_a + {1 \over 2!}(t^a{}_a t^b{}_b - t^a{}_b t^b{}_a) 
\ee
with $t^a{}_b = - i \kappa^2 \bar\psi^i \gamma^a \partial_b \psi^i$. 
The NLSUSY action (\ref{NLSUSYaction}) is invariant under NLSUSY transformations for $\psi^i$, 
\be
\delta_\zeta \psi^i = {1 \over \kappa} \zeta^i 
- i \kappa \bar\zeta^j \gamma^a \psi^j \partial_a \psi^i, 
\label{NLSUSY}
\ee
where $\zeta^i$ are the constant (Majorana) spinor parameters. 
The NLSUSY transformations (\ref{NLSUSY}) satisfy a closed off-shell commutator algebra, 
\be
[\delta_{\zeta_1}, \delta_{\zeta_2}] = \delta_P(\Xi^a), 
\label{NLSUSYcomm}
\ee
where $\delta_P(\Xi^a)$ means a translation with the parameters 
$\Xi^a = 2 i \bar\zeta_1^i \gamma^a \zeta_2^i$. 

Next we shall exhibit a (pure) SYM action for $N = 3$ SUSY in $d = 2$ in terms of a gauge supermultiplet 
corresponding to the helicity states for the irreducible representation of $SO(3)$ SP algebra, 
\be
\left[ \ \underline{1} (+1), 
\underline{3} \left( +{1 \over 2} \right), 
\underline{3} (0), 
\underline{1} \left( -{1 \over 2} \right) \ \right] 
+ [ {\rm CPT\ conjugate} ], 
\label{states}
\ee
where $\underline{n} (\lambda)$ means the dimension $\underline{n}$ 
and the helicity $\lambda$ of the irreducible representation. 
Component fields which belong to the adjoint representation of gauge group $G$ are defined as 
\be
V(x) = \{ v^a(x), \lambda^i(x), A^i(x), \chi(x), \phi(x), D^i(x) \} \ \ (i, j, \cdots = 1, 2, 3), 
\label{SYMcomp}
\ee
where we denote $v^a$ for vector fields, $\lambda^i$ and $\chi$ for (Majorana) spinor fields, 
$A^i$ for scalar fields, $\phi$ for pseudo scalar fields 
and $D^i$ for auxiliary scalar fields, respectively. 
These component fields are $V(x) = V^I(x) T^I$ with generators $T^I$ of $G$ 
satisfying $[T^I, T^J] = i f^{IJK} T^K$. 

The $N = 3$ (pure) SYM action in terms of the component fields (\ref{SYMcomp}) is given by 
\ba
S_{\rm SYM} \A = \A \int d^2x \ {\rm tr} \left\{ - {1 \over 4} (F_{ab})^2 
+ {i \over 2} \bar\lambda^i \!\!\not\!\!D \lambda^i 
+ {1 \over 2} (D_a A^i)^2 
+ {i \over 2} \bar\chi \!\!\not\!\!D \chi 
+ {1 \over 2} (D_a \phi)^2 + {1 \over 2} (D^i)^2 
\right. 
\nonu
\A \A 
- ig \{ \epsilon^{ijk} A^i \bar\lambda^j \lambda^k - [A^i, \bar\lambda^i] \chi 
+ \phi (\bar\lambda^i \gamma_5 \lambda^i + \bar\chi \gamma_5 \chi) \} 
\nonu
\A \A 
\left. 
+ {1 \over 4} g^2 ([A^i, A^j]^2 + 2 [A^i, \phi]^2) \right\}, 
\label{SYMaction}
\ea
where $g$ is the gauge coupling constant, 
$D_a$ and $F_{ab}$ are the covariant derivative and the non-Abelian gauge field strength defined as 
\ba
\A \A 
D_a \varphi = \partial_a \varphi - ig [v_a, \varphi], 
\nonu
\A \A 
F_{ab} = \partial_a v_b - \partial_b v_a - ig [v_a, v_b]. 
\ea
The SYM action (\ref{SYMaction}) is invariant under the following $N = 3$ LSUSY transformations 
parametrized by $\zeta^i$, 
\ba
\delta_\zeta v^a \A = \A i \bar\zeta^i \gamma^a \lambda^i, 
\nonu
\delta_\zeta \lambda^i 
\A = \A \epsilon^{ijk} (D^j - i \!\!\not\!\!D A^j) \zeta^k 
+ {1 \over 2} \epsilon^{ab} F_{ab} \gamma_5 \zeta^i 
- i \gamma_5 \!\!\not\!\!D \phi \zeta^i 
\nonu
\A \A 
+ ig ([A^i, A^j] \zeta^j + \epsilon^{ijk} [A^j, \phi] \gamma_5 \zeta^k), 
\nonu
\delta_\zeta A^i \A = \A \epsilon^{ijk} \bar\zeta^j \lambda^k - \bar\zeta^i \chi, 
\nonu
\delta_\zeta \chi 
\A = \A (D^i + i \!\!\not\!\!D A^i) \zeta^i 
+ ig (\epsilon^{ijk} A^i A^j \zeta^k - [A^i, \phi] \gamma_5 \zeta^i), 
\nonu
\delta_\zeta \phi \A = \A \bar\zeta^i \gamma_5 \lambda^i, 
\nonu
\delta_\zeta D^i 
\A = \A - i \epsilon^{ijk} \bar\zeta^j \!\!\not\!\!D \lambda^k 
- i \bar\zeta^i \!\!\not\!\!D \chi 
+ ig (\bar\zeta^i [\lambda^j, A^j] + \bar\zeta^j [\lambda^i, A^j] - \bar\zeta^j [\lambda^j, A^i] 
\nonu
\A \A 
- \epsilon^{ijk} \bar\zeta^j [\chi, A^k] 
+ \epsilon^{ijk} \bar\zeta^j \gamma_5 [\lambda^k, \phi] 
+ \bar\zeta^i \gamma_5 [\chi, \phi]), 
\ea
which satisfy the ordinary off-shell commutator algebra, 
\be
[\delta_{\zeta_1}, \delta_{\zeta_2}] = \delta_P(\Xi^a) + \delta_G(\theta) + \delta_g(\theta), 
\ee
where $\delta_G(\theta)$ means a commutator of the gauge transformation 
$\delta_G(\theta) \varphi = ig [\theta, \varphi]$ 
with a generator $\theta = - 2 (i \bar\zeta_1^i \gamma^a \zeta_2^i v_a 
- \epsilon^{ijk} \bar\zeta_1^i \zeta_2^j A^k - \bar\zeta_1^i \gamma_5 \zeta_2^i \phi)$, 
while $\delta_g(\theta)$ is the $U(1)$ gauge transformation only for $v^a$ with $\theta$. 

In order to see the relation between the NLSUSY action (\ref{NLSUSYaction}) 
and the SYM action (\ref{SYMaction}), we construct the SUSY invariant relations 
in which the component fields (\ref{SYMcomp}) are expressed in terms of $\psi^i$ ($i, j, \cdots = 1,2,3$) as 
%
$V(x) = V(\psi(x))$. 
%
As discussed in the previous works in the superfield formulation \cite{WB}, 
the hybrid transformations of L and NLSUSY transformations of $V(x)$ 
and the adoption of the subsequent SUSY invariant constraints produce systematically 
the SUSY invariant relations. 
They satisfy the commutator algebra (\ref{NLSUSYcomm}) 
as in the case of the Abelian gauge theories \cite{STT2,ST6-lin} 
and the NLSUSY GR \cite{KS1}, i.e. the SUSY transformations 
are the square root of the translation ($GL(4,R)$ in NLSUSY GR). 
The explicit form of the SUSY invariant relations for the $d = 2$, $N = 3$ gauge supermultiplet 
in the leading orders of $\kappa$ is 
\ba
v^{aI} 
\A = \A - {i \over 2} \kappa \epsilon^{ijk} \xi^{iI} 
\bar\psi^j \gamma^a \psi^k 
( 1 - i \kappa^2 \bar\psi^l \!\!\not\!\partial \psi^l ) 
+ {1 \over 4} \kappa^3 
\epsilon^{ab} \epsilon^{ijk} \xi^{iI} \partial_b 
( \bar\psi^j \gamma_5 \psi^k \bar\psi^l \psi^l ) 
+ {\cal O}(\kappa^5), 
\nonu
\lambda^{iI} 
\A = \A \epsilon^{ijk} \xi^{jI} \psi^k 
( 1 - i \kappa^2 \bar\psi^l \!\!\not\!\partial \psi^l ) 
\nonu
\A \A 
+ {i \over 2} \kappa^2 \xi^{jI} \partial_a 
\{ \epsilon^{ijk} \gamma^a \psi^k \bar\psi^l \psi^l 
+ \epsilon^{ab} \epsilon^{jkl} 
( \gamma_b \psi^i \bar\psi^k \gamma_5 \psi^l 
- \gamma_5 \psi^i \bar\psi^k \gamma_b \psi^l ) \} 
+ {\cal O}(\kappa^4), 
\nonu
A^{iI} 
\A = \A \kappa \left( {1 \over 2} \xi^{iI} \bar\psi^j \psi^j 
- \xi^{jI} \bar\psi^i \psi^j \right) 
( 1 - i \kappa^2 \bar\psi^k \!\!\not\!\partial \psi^k ) 
- {i \over 2} \kappa^3 \xi^{jI} \partial_a 
( \bar\psi^i \gamma^a \psi^j \bar\psi^k \psi^k ) 
+ {\cal O}(\kappa^5), 
\nonu
\chi^I 
\A = \A \xi^{iI} \psi^i 
( 1 - i \kappa^2 \bar\psi^j \!\!\not\!\partial \psi^j ) 
+ {i \over 2} \kappa^2 \xi^{iI} \partial_a 
( \gamma^a \psi^i \bar\psi^j \psi^j ) 
+ {\cal O}(\kappa^4), 
\nonu
\phi^I 
\A = \A - {1 \over 2} \kappa \epsilon^{ijk} \xi^{iI} \bar\psi^j \gamma_5 \psi^k 
( 1 - i \kappa^2 \bar\psi^l \!\!\not\!\partial \psi^l ) 
- {i \over 4} \kappa^3 
\epsilon^{ab} \epsilon^{ijk} \xi^{iI} \partial_a 
( \bar\psi^j \gamma_b \psi^k \bar\psi^l \psi^l ) 
+ {\cal O}(\kappa^5), 
\nonu
D^{iI} 
\A = \A 
{1 \over \kappa} \xi^{iI} \vert w \vert 
- i \kappa \xi^{jI} \partial_a \{ \bar\psi^i \gamma^a \psi^j 
( 1 - i \kappa^2 \bar\psi^k \!\!\not\!\partial \psi^k ) \} 
\nonu
\A \A 
- {1 \over 8} \kappa^3 
\Box \{ ( \xi^{iI} \bar\psi^j \psi^j - 4 \xi^{jI} \bar\psi^i \psi^j ) 
\bar\psi^k \psi^k \} 
+ {\cal O}(\kappa^5), 
\label{SYMinv}
\ea
where $\xi^{iI}$ are arbitrary real constants related to the constant terms 
of the auxiliary fields $D^{iI} = D^{iI}(\psi)$. 
The relations (\ref{SYMinv}) have the same form as those in Ref.\cite{ST4-lin} 
except the constants $\xi^{iI}$ with the indices of the gauge group $G$. 

By substituting Eqs.(\ref{SYMinv}) into the SYM action (\ref{SYMaction}), 
we can show it reduces to the NLSUSY action (\ref{NLSUSYaction}) for $N = 3$ SUSY as 
\be
S_{\rm SYM}(\psi) = - (\xi^{iI})^2 S_{\rm NLSUSY} + [{\rm surface\ terms}] 
\ee
at least in the leading orders. 
Indeed, the kinetic terms (with the $(D^{iI})^2$-terms) in $S_{\rm SYM}$ reduces to $S_{\rm NLSUSY}$ as 
\ba
S^{\rm kin.}_{\rm SYM}(\psi) 
\A = \A \int d^2x \ {\rm tr} \left\{ - {1 \over 4} (f_{ab})^2 
+ {i \over 2} \bar\lambda^i \!\!\not\!\partial \lambda^i 
+ {1 \over 2} (\partial_a A^i)^2 
+ {i \over 2} \bar\chi \!\!\not\!\partial \chi 
+ {1 \over 2} (\partial_a \phi)^2 + {1 \over 2} (D^i)^2 \right\} 
\nonu
\A = \A - (\xi^{iI})^2 S_{\rm NLSUSY} 
\label{kin-NLSUSY}
\ea
(at least up to ${\cal O}(\kappa^2)$), where $f_{ab} = \partial_a v_b - \partial_b v_a$. 
On the other hand, each interaction terms at ${\cal O}(g)$ in $S_{\rm SYM}$ 
give the following (non-vanishing) four and/or six superon (NG-fermion) self-interaction terms 
up to ${\cal O}(\kappa^3)$, 
\ba
\A \A {\rm (a)}\ g \kappa \epsilon^{ijk} f^{IJK} \xi^{iI} \xi^{jJ} \xi^{kK} 
\bar\psi^l \psi^l \bar\psi^m \psi^m\ ({\rm at}\ {\cal O}(\kappa)), \nonu
\A \A {\rm (b)}\ i g \kappa^3 \epsilon^{ijk} f^{IJK} \xi^{iI} \xi^{jJ} \xi^{kK} 
\bar\psi^l \psi^l \bar\psi^m \psi^m \bar\psi^n \!\!\not\!\partial \psi^n\ ({\rm at}\ {\cal O}(\kappa^3)). 
\label{selfint}
\ea
However, they cancel with each other in $S_{\rm SYM}$, i.e. the interaction terms in $S_{\rm SYM}$ vanish as 
\ba
S^{\rm int.}_{\rm SYM}(\psi)\ {\rm at}\ {\cal O}(g) 
\A = \A \int d^2x \ (-ig) \ {\rm tr} \bigg\{ - f_{ab} v^a v^b 
+ \partial_a A^i [v^a, A^i] + \partial_a \phi [v^a, \phi] 
\nonu
\A \A 
+ {i \over 2} (\bar\lambda^i \gamma^a [v_a, \lambda^i] 
+ \bar\chi \gamma^a [v_a, \chi]) 
\nonu
\A \A 
+ \epsilon^{ijk} A^i \bar\lambda^j \lambda^k - [A^i, \bar\lambda^i] \chi 
+ \phi (\bar\lambda^i \gamma_5 \lambda^i + \bar\chi \gamma_5 \chi) \bigg\} 
\nonu
\A = \A 0 
\label{int-NLSUSY}
\ea
(at least up to ${\cal O}(\kappa^3)$). 
Each interaction (potential) terms at ${\cal O}(g^2)$ in the SYM action (\ref{SYMaction}) 
vanishes in terms of $\psi^i$ due to $(\psi^i)^7 \equiv 0$ for $d = 2$, $N = 3$ SUSY. 
From Eqs.(\ref{kin-NLSUSY}) and (\ref{int-NLSUSY}) we conclude that the NLSUSY action (\ref{NLSUSYaction}) 
for $N = 3$ SUSY is related to the $N = 3$ SYM action (\ref{SYMaction}) as 
\be
- (\xi^{iI})^2 S_{\rm NLSUSY} = S_{\rm SYM} + [{\rm surface\ terms}]. 
\label{NL/LSUSY}
\ee
(From the previous works we anticipate that the relation (\ref{NL/LSUSY}) holds in all orders of $\psi^i$, 
though yet to be confirmed.) 
For the case of $d = 2$, $N = 2$ SUSY, each interaction terms at ${\cal O}(g)$ in a pure SYM action 
vanishes in terms of $\psi^i$, which means that the relation (\ref{NL/LSUSY}) is trivial 
for the $d = 2$, $N = 2$ pure SYM theory. 

Let us summarize our results as follows. 
In this letter we explicitly discuss for $N = 3$ SUSY in $d = 2$ 
the relation between the NLSUSY action (\ref{NLSUSYaction}) and the SYM action (\ref{SYMaction}). 
The SUSY invariant relations (\ref{SYMinv}) are constructed by extending constant terms 
of the auxiliary fields to $\xi^{iI}$ ($i, j, \cdots = 1, 2, 3$ 
and $I, J, \cdots = 1, 2, \cdots, {\rm dim}G$) with the indices of the gauge group $G$ 
in $D^{iI} = D^{iI}(\psi)$. By substituting SUSY invariant relations into the SYM action 
we show the NL/L SUSY relation (\ref{NL/LSUSY}) for the $d = 2$, $N = 3$ SYM theory, 
in which the interaction terms vanish as (\ref{int-NLSUSY}) 
by means of the cancellations among the superon (NG-fermion) self-interaction terms (\ref{selfint}). 
We anticipate similar results for the (pure) SYM theory in $d = 4$. 
The further investigations for the SYM theory with matter supermultiplet (the SUSY QCD theory) 
in NL/L SUSY relation are interesting and crucial.

%
%

\newpage

%
\newcommand{\NP}[1]{{\it Nucl.\ Phys.\ }{\bf #1}}
\newcommand{\PL}[1]{{\it Phys.\ Lett.\ }{\bf #1}}
\newcommand{\CMP}[1]{{\it Commun.\ Math.\ Phys.\ }{\bf #1}}
\newcommand{\MPL}[1]{{\it Mod.\ Phys.\ Lett.\ }{\bf #1}}
\newcommand{\IJMP}[1]{{\it Int.\ J. Mod.\ Phys.\ }{\bf #1}}
\newcommand{\PR}[1]{{\it Phys.\ Rev.\ }{\bf #1}}
\newcommand{\PRL}[1]{{\it Phys.\ Rev.\ Lett.\ }{\bf #1}}
\newcommand{\PTP}[1]{{\it Prog.\ Theor.\ Phys.\ }{\bf #1}}
\newcommand{\PTPS}[1]{{\it Prog.\ Theor.\ Phys.\ Suppl.\ }{\bf #1}}
\newcommand{\AP}[1]{{\it Ann.\ Phys.\ }{\bf #1}}

\end{document}